\documentclass[twocolumn,aps,prd,groupedaddress,showpacs]{revtex4-1}
\usepackage{graphicx}
\usepackage{bm}
\usepackage{epsfig}
\usepackage{color}
\usepackage{enumerate}
\usepackage{ulem}

\begin{document}

\title{Simulation studies of dark energy clustering induced by the formation of dark matter halos}
\author{Qiao Wang}
\author{Zuhui Fan}
\affiliation{Department of Astronomy, School of Physics, Peking University, Beijing, 100871, People's Republic of China}
   
\date{\today}

\begin{abstract}
In this paper, we present a simulation method within the two-component spherical collapse model to 
investigate dark energy perturbations associated with the formation of dark matter halos.
The realistic mass accretion history of a dark matter halo
taking into account its fast and slow growth is considered by imposing suitable initial conditions 
and isotropized virializations for the spherical collapse process. The dark energy component is 
treated as a perfect fluid described by two important parameters, 
the equation of state parameter $w$ and the sound speed $c_s$. Quintessence models with $w>-1$ are analyzed. 
We adopt the Newtonian gauge to describe the 
spacetime which is perturbed mainly by the formation of a dark matter halo. 
It is found that the dark energy density perturbation $\delta_{DE}$ depends on  
$w$ and $c_s$, and its behavior follows closely the gravitational potential $\Phi$ of the dark matter halo with  
$\delta_{DE}\approx -(1+w)\Phi/c_s^2$. For $w>-1$, the dark energy perturbation presents a clustering behavior 
with $\delta_{DE}>0$ during the entire formation of the dark matter halo, from linear to nonlinear and virialized stages. 
The value of $\delta_{DE}$ increases with the increase of the halo mass.
For a cluster of mass $M\sim 10^{15} M_{\odot}$, $\delta_{DE}\sim 10^{-5}$ within the virialized region for 
$c_s^2 \in [0.5, 1]$, and it can reach $\delta_{DE}=O(1)$ with $c_s^2=0.00001$.
For a scalar-field dark energy model, 
we find that with suitably modeled $w$ and $c_s$, its perturbation behavior associated with the nonlinear formation 
of dark matter halos can well be analyzed using the fluid approach, demonstrating the validity of the fluid description
for dark energy even considering its perturbation in the stage of nonlinear dark matter structure formation.

\end{abstract}


\maketitle

\section{Introduction}

Since the discovery of the accelerating expansion of the universe, tremendous progress 
has been achieved in dark energy studies\cite{Amanullah10}\cite{Percival10}\cite{Reid10}\cite{Larson11}\cite{Copeland06}\cite{Frieman08}.
Future generations of cosmological observations are expected to be able to provide us tight 
constraints on properties of dark energy \cite{Zhan06}\cite{Barnard08}\cite{Zhan09}\cite{Zhao10}\cite{Mortonson10}. The full realization of their constraining power, however, depends on our thorough understandings on how different dark energy models affect the expansion of the universe and the formation and evolution of cosmic structures differently 
\cite{Ma99}\cite{Klpyin03}. 

For dynamical dark energy models, apart from the background field that drives the accelerating 
expansion of the universe, dark energy perturbations exist intrinsically. 
Their behaviors on large scales and the effects on the linear power spectrum of the matter
density perturbations and the anisotropy of the cosmic microwave background radiation (CMB) have been 
extensively investigated and taken into account in deriving cosmological constraints from CMB observations 
\cite{GZhao05}\cite{Li08b}\cite{Komatsu11}\cite{Fang08}.
On small scales of the nonlinear structure formation, they should also be considered.
Therefore a fully consistent simulation to study the impacts of dark energy on the structure formation
should simulate the evolution of the dark energy component explicitly
together with the matter components (dark matter and baryonic matter). 
The additional dark energy component can however complicate the
numerical simulation significantly. Therefore in many studies, the dark energy
is not included as an independent component in simulations. Instead, approximations or 
simplifications suitable to specific models are often adopted so that the effects of dark energy perturbations
on the structure formation can be partially taken into account \cite{Maccio04}\cite{Baldi09}\cite{Baldi10a}\cite{Baldi10b}\cite{Baldi11}\cite{Alimi10}. 
In a very recent series of papers regarding coupled dark matter-dark energy models, 
$f(R)$ gravity models and scalar-tensor theories, the corresponding scalar field component has been 
explicitly included in simulations \cite{BLi09}\cite{BLi11a}\cite{BLi11b}\cite{BLi11c}\cite{GZhao11}. 
Such simulations not only can study the structure formation self-consistently,
but also can provide the detailed dynamical evolution of the relevant scalar field component, which 
may contain additional information in differentiating different models. 

To study the effects of dark energy and particularly the behavior of dark energy perturbations
along with the formation of nonlinear structures, the analyses under the spherical collapse model
are extensively performed. Although it is a simplified model comparing with the full cosmological simulations, 
it allows us to isolate different effects and to explore a broad parameter space efficiently.
Studies along this line have shown that the simple extrapolation from the amplitude of 
the dark energy perturbation obtained in the linear stage of structure formation can grossly overestimate 
the level of the dark energy perturbation in the nonlinear epoch of the structure formation 
 \cite{Dutta07}\cite{Mota08}\cite{Wang09}.
For minimally-coupled quintessence dark energy models with $c_s^2\sim 1$, 
the dark energy perturbation induced by the formation of cluster-scale dark matter halos
remains at the level of $< 10^{-5}$, in contrast with the amplitude of $\sim 10^{-2}$ obtained by
extrapolating the linear analyses to late stages of dark matter halo formation where nonlinear
processes actually happen\cite{Dutta07}\cite{Mota08}\cite{Wang09}\cite{Maor05}\cite{Nunes06}\cite{Abramo07}\cite{Basse11}\cite{Sefusatti11}\cite{Devi11}\cite{He09}\cite{Creminell10}\cite{He10}.    
Such a large difference indicates that the virialization of dark matter halos plays 
an important role in determining the behavior of dark energy perturbations in the halo region.
Previous analyses take into account the virialization 
process by adding in a simple prescription into the analytical formulation 
to prevent the halo material from collapsing all the way to the center (e.g., \cite{Mota08}\cite{Wang09}). Although
such a modeling can catch the basics of the virialization process, it cannot
describe it very realistically. Particularly, it cannot handle properly the shell crossing 
that is important for the virialization process. 

To overcome the shortcomings of the artificial treatment of the virialization process, 
in this paper, we carry out numerical simulations for a set of two-component 
spherical systems. Following Lu et al. \cite{Lu06}, by specifying proper initial conditions,
the realistic mass assembling history for a dark matter halo is considered in the 
spherical 1-D simulations. 
The tangential velocity is also taken into account,  which can affect the final density 
profile of a dark matter halo significantly. With these implementations, the formation of 
dark matter halos can be suitably simulated, and the corresponding dark energy perturbations 
can be investigated. In our simulations, the dark energy component is regarded as an ideal fluid
specified by two important parameters, the equation of state parameter $w$ and the
sound speed parameter $c_s$. We investigate the dependence of dark energy perturbations
on various quantities, such as $w$, $c_s$ and the mass of the dark matter halo. The
validity of the fluid approach taking into account the dark energy perturbations
is also analyzed.

The rest of the paper is organized as follows. In Sec.~\ref{sec:formula}, we present the formulations related
to the two-component spherical system. Sec.~\ref{sec:simulation} describes the methodology of simulations. 
The results and discussions are shown in Sec.~\ref{sec:result} and Sec.~\ref{sec:discussion}, respectively. 

\section{Formulations\label{sec:formula}}

In this study, we simulate a set of two-component spherical systems, where the dark matter and dark energy
components are noncoupling except gravitational interactions. Specifically, we follow the 
evolution of a spherically overdense region of dark matter, and analyze the induced dark energy perturbations.
Because of the small amplitudes, we assume that dark energy perturbations have no effects on 
the formation of dark matter halos \cite[e.g.,][]{Wang09}.  
The detailed methodology for simulating the dark matter halo formation will be described in 
next section. Here we present the formulations used in calculating dark energy perturbations
associated with the formation of dark matter halos.

To consider simulations with shell crossing, it is convenient to adopt the Newtonian gauge to 
describe the perturbed spacetime metric, which has been shown to be valid even in the stage of 
nonlinear halo formation \cite{vanAcoleyen08}. The metric can be written as follows, 
\begin{equation}
 ds^2 =   -(1+2\Phi)d t^2 + a^2 (1-2\Psi)dx_i dx^i,
\label{eq:le}
\end{equation}
where $a$ is the cosmic scale factor, and $\Phi$ and $\Psi$ are the Newtonian potential and 
the spatial curvature, respectively. The flat universe is considered here. In the case of spherical
symmetry, $dx_i dx^i=dr^2+r^2d\Omega$.
In Fourier space, the Einstein's equations in terms of the comoving coordinates then read 
\begin{equation}
k ^2 \Psi + 3 a^2 H [\dot{\Psi} + H\Phi] = -4\pi G a^2 \sum_i \delta\rho_i, \\ \nonumber
\end{equation}
\begin{equation}
k ^2 [\Psi - \Phi] = 12\pi G a^2 \sum_i (\bar{\rho}_i+\bar{p}_i)\sigma_i,\\ \nonumber
\end{equation}
where $H=\dot a/a$ is the Hubble parameter, and the dot represents the derivative with respect to time $t$.
The sum is over different components with $i=DM$ for dark matter and $i=DE$ for dark energy.  
The quantities $\bar{\rho}_i$ and $\bar{p}_i$ are the
background energy density and pressure for component $i$, respectively,
$\delta \rho_i \equiv (\rho_i - \bar{\rho}_i)$,  
and $\sigma_i$ is the anisotropic stress perturbation for component $i$ \cite{Ma95}. 
For the dark energy component, it in general can have the anisotropic stress perturbation 
which can be significant if it is a relativistic fluid (e.g., \cite{Mota07}\cite{Appleby10}\cite{Calabrese11}). 
On the other hand, cosmological observations can only put very weak constraints on it, and
thus it may be phenomenologically sufficient for broad classes of dark energy models
to regard the dark energy component as a perfect fluid \cite{Mota07}.  
Furthermore, it has been shown in \cite{Mainini08} that for self-interacting scalar-field dark energy models,
the anisotropic stress perturbation vanishes and thus they can be described as perfect fluids.   
The equivalence between a scalar field model and the perfect-fluid description is also 
demonstrated by our simulation studies shown here in \S IV.D.
Therefore in our analyses, we treat both the dark matter and the dark energy components as perfect fluids
with $\sigma_i=0$. We then have $\Phi = \Psi$. 
In the following, $\Psi$ will be substituted by $\Phi$. 

On scales of dark matter halos, we have $k^2 \Phi \gg H(\dot{\Phi} + H \Phi )$. Furthermore, 
$\delta \rho_{DE} \ll \delta \rho_{DM}$. Then Eq. (2) reduces to  
\begin{equation}
k^2 \Phi = -4 \pi G a^2 \delta \rho_{DM}.
\label{eq:Ng}
\end{equation}
Therefore the potential is determined by the dark matter distribution only. 
This simplifies the calculations significantly.  

For the dark energy component, we define $\delta_{DE} \equiv \delta \rho _{DE} / \bar{\rho} _{DE} $ and 
$\theta_{DE} \equiv i \vec k \cdot \vec v_{DE}$ with $\vec v_{DE}$ being the velocity of the dark energy fluid. 
Because of the smallness of their amplitudes, we only analyze the linearized dynamical equations for dark energy perturbations, which are
given by 
\begin{equation}
\dot{\delta}_{DE} + 3 H\left( c_s^2 - w \right) \delta_{DE} + (1+ w ) \left( \frac{\theta_{DE}}{a} - 3 \dot{\Phi} \right)=0,
\label{eq:dlt}
\end{equation}
\begin{equation}
\dot{\theta}_{DE} + \left[ H(1-3w)+\frac{\dot{w}}{1+w} \right] \theta_{DE} \\ 
-\frac{k^2}{a}\left(\frac{c_s^2}{1+w} \delta_{DE} + \Phi\right)=0,
\label{eq:the}
\end{equation}
where $w \equiv \bar{p}_{DE} / \bar{\rho}_{DE}$ is the equation-of-state parameter of the dark energy fluid and 
$c_s$ corresponds to the sound speed parameter in the rest frame of the dark energy fluid and is defined as 
$c_s^2\equiv \delta p_{DE}/ \delta \rho_{DE}$.
If the perturbation is pure adiabatic, we have $c_s^2<0$ for the dark energy fluid with $w<0$,
then the perturbation is unstable (e.g., \cite{Mainini08}\cite{Langlois07}). Therefore from physical considerations, the
perturbation of the dark energy fluid cannot be pure adiabatic. 
Here we regard $c_s^2$ as a parameter \cite{DeDeo03}\cite{Bean04}\cite{Hannestad05}\cite{Xia08}\cite{Ballesteros10}, and analyze
how the behavior of dark energy perturbations depends on it.

The above equations form the bases for simulating dark energy perturbations.
Specifically, the formation of a spherical dark matter halo is followed by numerical simulations
taking into account the effect of background dark energy but without including dark energy perturbations.
We then calculate in Fourier space the potential by Eq.~(\ref{eq:Ng}), and further dark energy perturbations
by Eq.~(\ref{eq:dlt}) and Eq.(\ref{eq:the}). Finally, dark energy perturbations in real space are 
computed by inverse Fourier transformations. 

In the context of scalar-field dark energy models, their fluid description including the presence of perturbations
relies on finding suitable correspondences between $w$ and $c_s^2$ and the scalar field \cite{Mainini08}.
For a scalar field $\phi$ with potential $V(\phi)$, we have $w=\bar p_{DE}/\bar \rho_{DE}$ where
\begin{equation}
\bar \rho_{DE}={1\over 2}{\dot\phi}^2+V(\phi), \quad \bar p_{DE}={1\over 2}{\dot\phi}^2-V(\phi). 
\label{eq:rhop}
\end{equation}

The dynamical evolution of the field is given by 
\begin{equation}
\ddot{\phi}+3H\dot{\phi}+V'=0, 
\label{eq:bgsf}
\end{equation}
where $V'=dV/d\phi$.

The linear perturbation equation is then
\begin{equation}
\ddot {\delta \phi }+ 3H \dot{\delta \phi}+\frac{k^2}{a^2} \delta \phi + \delta V' = 4\dot{\Phi}\dot{\phi}.
\label{eq:sf}
\end{equation}
The corresponding density and pressure perturbations in the fluid description are given by \cite{Mainini08}
\begin{equation}
\delta \rho_{DE} = - \dot{\phi}^2\Phi + \dot{\phi}\dot{\delta \phi} + \delta V,
\label{eq:dltsf}
\end{equation}
and
\begin{equation}
\delta p_{DE} = - \dot{\phi}^2\Phi + \dot{\phi}\dot{\delta \phi} - \delta V=\delta \rho_{DE}-2\delta V=\delta \rho_{DE}-2 V'\delta \phi.
\end{equation}
The divergence of the velocity $\theta_{DE}$ corresponds to \cite{Mainini08}
\begin{equation}
\theta_{DE} \equiv \frac{k^2}{a\dot{\phi}} \delta \phi. 
\label{eq:the_sf}
\end{equation}
We then obtain the sound speed-related parameter $c_s^2$ by \cite{GZhao05}
\begin{equation}
c_s^2 =\delta p_{DE}/\delta\rho_{DE}= 1+\frac{a \theta_{DE} }{ k^2 \delta_{DE} }\left[ 3H(1-w^2) + \dot{w}\right].
\label{eq:wp}
\end{equation}
It is noted that for a given scalar field model, $c_s^2$ and $w$ are related to each other. 
On the other hand, in the general fluid description without concerning a particular underlying scalar field model,
$c_s^2$ and $w$ can be regarded as two independent parameters that are used to describe various models.
 
To test the validity of the fluid description including dark energy perturbations,
for a specific scalar field model with a double-exponential potential given by
\begin{equation}
V(\phi) = V_0(e^{\alpha \sqrt{8\pi G}\phi}+e^{\beta \sqrt{8\pi G}\phi})
\label{eq:poten}
\end{equation}
where $\alpha=6$ and $\beta=0.1$ (e.g., \cite{Wang09}), we simulate the perturbations both for the scalar field explicitly by Eq.~(\ref{eq:sf}) and with the 
fluid approach with the corresponding $w$ given by Eq.~(\ref{eq:rhop}) and $c_s^2$ given by  Eq.~(\ref{eq:wp}).
The results are then compared. 

\section{Simulation Algorithm\label{sec:simulation}}

As discussed in Sec.II, we ignore the feedback effects of dark energy perturbations on the 
dynamical evolution of the dark matter component. Therefore in our studies, 
the formation and evolution of a dark matter halo are simulated independently of dark energy perturbations. 
Specifically, the dark matter mass of the simulated halo is divided into equal-mass shells.
These shells are very much analogous to mass particles in N-body simulations, and we refer them as shell-particles.   
Their dynamical motions are followed using a one-dimensional code adapted from 
Lu et al. \cite{Lu06} taking into account the effect of the background dark energy component. 
To compute the dark energy perturbations induced by the gravitational potential of the dark matter halo, 
at each time step, from the positions of the shell-particles, we construct the spatial dark matter mass density
field on regular radial-mesh bins with equal width in the radial coordinate $r$.
This spherical dark matter density field is then transformed into the Fourier space, 
and the corresponding gravitational potential is obtained by Eq~(\ref{eq:Ng}).
For each Fourier mode, we calculate the dark energy perturbation with Eq.~(\ref{eq:dlt}) and Eq.~(\ref{eq:the}).
With the inverse Fourier transformation, we then obtain the dark energy perturbation in the configuration space.

\subsection{Simulations for the formation of spherical dark matter halos}

Although highly simplified, the spherical collapse model (SCM) catches the important aspects of halo
formation, such as giving rise to an important collapse threshold that can be used 
to quantify statistically the formation for dark matter halos. On the other hand, the simple top-hat SCM  
cannot describe realistically the mass assembly of dark matter halos, which involves the 
early-stage fast accretion and merging and late-stage slow growth (e.g., \cite{Zhao03a}).    
Furthermore, the analytical treatment of SCM cannot model the process of virialization of dark matter halos
naturally. Lu et al. \cite{Lu06} develop a one-dimensional numerical simulation method\cite{Shapiro04}\cite{Sanch06} 
to simulate the formation of a spherical dark matter halo. By suitably
choosing the initial dark matter mass distribution, the fast- and slow-accretions of the 
halo formation can be properly taken into account. Here we follow this strategy to simulate the
formation of dark matter halos with the important modification to incorporate the effect of 
the background dark energy.  

In the 1-D simulations, the mass of the dark matter component is assigned to equal-mass shells, which are analogous
to particles in N-body simulations. The acceleration $a_r$ of a shell-particle at position $r$ depends on the 
matter content within $r$. Specifically, we have

\begin{equation}
a_r =  -\frac{H^2}{2} \Omega_{DE} (t)(1+3w) r- \frac{G M(r) r}{(r ^2 + \alpha_s^2)^{3/2}}+\frac{J^2}{r^3}.
\label{eq:dy}
\end{equation}
where the first term represents the contribution from the background dark energy with 
the equation of state parameter $w$ and the energy density parameter $\Omega_{DE} (t)$ in unit
of the critical density of the universe at time $t$.
Note that in general $w$ can be time dependent. 
The second term is the gravity from the dark matter component, and the third term is the centrifugal force from the 
angular momentum $J$. Both the Hubble parameter $H$ and the dark energy density parameter $\Omega_{DE}$
are time evolving and are dependent on the equation of state parameter $w$. 
In the second term, $M(r)=\sum_{r'<r} m_{r'}$ is the total mass within $r$ calculated
by summing over the mass of the shell-particles inside $r$. The softening parameter 
$\alpha_s$ is taken to be $\alpha_s=1\hbox{ kpc}$, which is about $0.0005R_{vir}$ for the virial radius
$R_{vir}\sim 2\hbox{Mpc}$ for clusters of galaxies \cite{Lu06}. 
The angular momentum $J$ is added in by hand to prevent an over-concentrated dark matter density profile.
Following Lu et al. \cite{Lu06}, a tangential velocity for a shell-particle is added in when it falls back to 
one half of its turn-around radius $R_t$ with the tangential and radial velocity dispersions,
$\sigma_t^2$ and $\sigma_r^2$, satisfying the relation
\begin{equation}
\frac{\sigma_t^2}{\sigma_r^2}=\frac{2}{1+(R_t/r_a)^{\beta}}.
\label{eq:part}
\end{equation}
where $\beta$ is taken to be $\beta=2$ following Lu et al. \cite{Lu06} who show that  
the results are insensitive to the specific value of $\beta$. 
The parameter $r_a$ is chosen to be the virial radius of the 
halo at time $a_c$, the transition time between the fast-accretion and slow-accretion phases 
[see Eq.~(\ref{eq:mah})] \cite{Lu06}. Specifically, the radial and tangential velocities are randomly generated
according to Gaussian distributions with the dispersions of $\sigma_r$ and $\sigma_t$, respectively.
Then the total kinetic energy of the shell-particle is partitioned into radial and tangential components according to 
the square of the ratio of the two random numbers. Note that we include the 
tangential velocity merely in Eq.~(\ref{eq:dy}) to modify the radial motion of a shell-particle
without really simulating its tangential motion. 

In our simulations, we use $10^5$ dark matter shell-particles. Further increasing the number 
does not change the results significantly. The symplectic integrator is employed \cite{Quinn97}.
The time step is chosen to be smaller than the minimum of the dynamical time scales
of the shell-particles.

For dark energy perturbations, we do not calculate them at the positions of the moving dark matter shell-particles.
Instead, they are done on a fixed regular radial-mesh of equal width along
the radial coordinate $r$ in accord with the Newtonian-gauge metric shown in Eq.~(\ref{eq:le}).  
At each time step of the simulation, we construct the spherical dark matter density field on that
radial-mesh from the positions of dark matter shell-particles.
Then the over-density field $\delta \rho_{DM}$ is obtained and transformed
into the Fourier space. The gravitational potential $\Phi (k)$ is calculated 
and the dark energy perturbation for each $k$ mode is further solved with Eq.~(\ref{eq:dlt}) and Eq.~(\ref{eq:the}).
Finally, the dark energy perturbation at a desired output time in the configuration space is obtained by 
the inverse Fourier transformation.

\subsection{Initial conditions and virialization}
Apart from the necessity to introduce tangential motions, cosmological simulation studies 
have shown that the universal density profile for dark matter halos
also depends sensitively on their mass assembling history, which cannot be described well
by the simple top-hat spherical collapse model \cite[e.g.,][]{Zhao03a}.
However, as demonstrated by Lu et al. \cite{Lu06},
this can be remedied within the spherical collapse framework by setting up properly the initial mass distribution 
for the dark matter component (instead of the simple top-hat density profile)
according to an approximate description about the realistic mass accretion process.

Considering only the growth of the amount of mass for a dark matter halo without
concerning the detailed processes, Wechsler et al \cite{Wechsler02} present an approximate
form for the time-dependence of its mass $M(a)$, which is given by
\begin{equation}
M(a)=M_0 \exp \left[ -a_c S\left(\frac{a_0}{a}-1\right)\right],
\label{eq:mah}
\end{equation}  
where $M_0$ is the halo mass at the observed epoch $a_0$, and $a_c$ is the characteristic scale factor
to divide the fast- and slow-accretions specified by $(d\ln M/d\ln a)|_{a=a_c}=a_0S$ and $S=2$.  
We can see that $a_c$ plays the critical role in describing the mass growth for a halo, which in turn
affects the final density profile, such as the concentration parameter of the NFW halos \cite{Navarro96}\cite{Navarro97}.
In other words, given a desired concentration parameter for the final density profile of a dark matter
halo, one can find a suitable value of $a_c$ so that Eq.(\ref{eq:mah}) can describe the mass assembly of the 
halo properly \cite{Lu06}.

In $\Lambda$CDM models, the empirical relation between the concentration parameter and the mass
of a halo has been investigated extensively from numerical simulations 
(e.g., \cite{Bullock01}\cite{Wechsler02}\cite{Zhao03a}\cite{Zhao03b}\cite{Zhao09}).
For example, \cite{Bullock01} presents a simple mass-concentration relation given by   
\begin{equation}
c_{vir} ^{\Lambda CDM}(M,z) \simeq \frac{9}{1+z}\left(\frac{M_0}{M_{*}}\right)^{-0.13},
\label{eq:concen}
\end{equation}
where $c_{vir}=R_{vir}/r_s$ with $r_s$ being the characteristic scale of an NFW halo, 
$M_0$ is the halo mass at the observed redshift $z$, and $M_{*} = 1.5 \times 10^{13} M_{\odot}$ derived 
from simulations \cite{Bullock01}.
Thus given a halo mass $M_0$, we can calculate the expected concentration parameter from Eq.(\ref{eq:concen}). 
We take $a_0=1$, i.e., $z=0$ as the final epoch. For $M_0=10^{15}\hbox{ M}_{\odot}$, 
$10^{14}\hbox{ M}_{\odot}$, and $10^{13}\hbox{ M}_{\odot}$, 
we have $c_{vir}^{\Lambda CDM} \sim 5.2$, $7.0$ and $9.5$, respectively. 
For dynamical dark energy models, the formation and evolution of dark matter halos 
can be different from those in $\Lambda$CDM models. Studies show that 
the differences can largely be attributed to the differences of the linear growth factor $D(z)$
of the dark matter density perturbations through its effect on the halo formation epoch (e.g., \cite{Dolag04}\cite{Grossi10}).
It is found that the concentration parameter $c_{vir}$ for a dynamical dark energy model can be
described well by the relation 
\begin{equation}
c_{vir} \approx c_{vir} ^{\Lambda CDM} D(z \rightarrow +\infty)/D^{\Lambda CDM}(z \rightarrow +\infty),
\label{eq:concende}
\end{equation}
where $D(z \rightarrow +\infty)$ and $D^{\Lambda CDM}(z \rightarrow +\infty)$ are the linear growth factor
evaluated at very high redshift $z\rightarrow \infty$ for the concerned model and for the $\Lambda$CDM model, 
respectively (\cite{Dolag04}\cite{Grossi10}). Here we adopt this relation to calculate the concentration parameter
for a particular dark energy model with $c_{vir} ^{\Lambda CDM}$ determined by Eq.(\ref{eq:concen}).
The linear growth factor is calculated by (e.g., \cite{Wang98})
\begin{equation}
\ddot{D}+2H\dot{D}-\frac{3}{2}H^2\Omega_{DM}(t)D=0,
\label{eq:lg}
\end{equation}
where the initial conditions at very high redshift are taken to be $D= C\times a$, and
$dD/dt=C\times da/dt$ with the constant $C$ determined by the normalization condition $D(z=0)=1$.
With the obtained $c_{vir}$, the corresponding parameter $a_c/a_0$ is found from Figure 4 of Lu et al. \cite{Lu06}.
We list them in Table~\ref{tab:sim}.

Based on Lu et al. \cite{Lu06}, we use the following methodology to find the initial dark matter mass distribution 
that can give rise to the mass accretion shown in Eq.~(\ref{eq:mah}). For a halo with 
mass $M_0$ at $a_0$, we assume that $M(a)$ in Eq.~(\ref{eq:mah}) is the mass within $M_0$ that is
virialized at the epoch $a$. According to the spherical collapse model, the linear extrapolated 
density perturbation averaged over the scale corresponding to $M(a)$ should reach the collapse threshold 
$\delta_c$ at time $a$. Then the corresponding perturbation at the initial epoch $a_i$ can be written as
\begin{equation}
\delta_i(M)=\delta_c {D(a_i)\over D[a(M)]},
\label{eq:di}
\end{equation}
where $a(M)$ is the inverse function of Eq.~(\ref{eq:mah}), representing the epoch 
when the amount of mass $M$ within the halo is virialized, and $\delta_i(M)$ is the 
initial value of the density perturbation averaged over the scale within which the contained mass is $M$. 
The subscript $i$ denotes the quantities at the initial epoch.
The initial radius of the region with the average density perturbation $\delta_i(M)$ is then given by 
\begin{equation}
r_i(M)=\left[ \frac{3M}{4 \pi \bar \rho_{DM} (a_i)[1+\delta_i(M)]} \right] ^{1/3},
\label{init_a}
\end{equation}
where $\bar \rho_{DM} (a_i)$ is the background dark matter density at $a_i$. 
The initial radial velocity is set by 
\begin{equation}
v_i(M)=H_ir_i - \frac{1}{3} H_i \left[ \frac{d \ln \delta}{d \ln a}\right]_{a_i}  \delta_i r_i, 
\label{init_b}
\end{equation}
where  the second term is the peculiar velocity due to the mass density perturbation 
\cite{Peebles80}\cite{Padmanabhan}.

We choose the initial redshift $z_i=30$  (correspondingly $a_i=a_0/31$), 
and the initial dark energy perturbations are set to be zero.  The cosmological parameters
are $\Omega_{DM0}=0.3$, $\Omega_{DE0}=0.7$, and $H_0=70\hbox{km/s/Mpc}$, where
$\Omega_{DM0}$ and $\Omega_{DE0}$ are the current dark matter and dark energy density parameters of the universe
in unit of the critical density. The value of $\delta_c$ is set to be $1.686$. 

\begin{figure}[ht]
\includegraphics[width=0.48\textwidth]{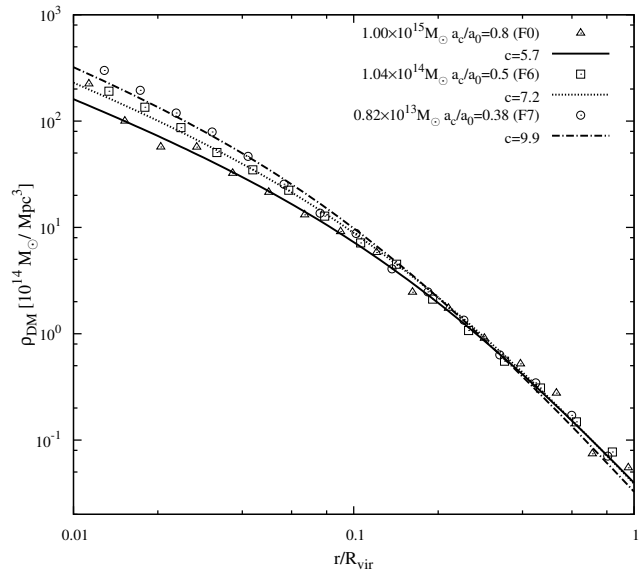}
\caption{\label{fig:PRO}Dark matter halo density profiles.  
The symbols are the results from simulations for models of F0 (triangles),
F6 (squares) and F7 (circles), respectively. The lines are the corresponding fitted NFW profiles
with the concentration parameters shown in the plot. }
\end{figure}

In Figure~\ref{fig:PRO}, we show the density profiles from our 1-D simulations
for three different halos. The virialized masses at $z=0$ from the simulations are
$M_0=1.00\times 10^{15}\hbox{ M}_{\odot}$, $1.04\times 10^{14}\hbox{ M}_{\odot}$, and $0.82\times 10^{13}\hbox{ M}_{\odot}$, 
respectively, in excellent accordance with $M_0=10^{15}\hbox{ M}_{\odot}$,$10^{14}\hbox{ M}_{\odot}$, 
and $10^{13}\hbox{ M}_{\odot}$, the chosen masses of the halos used to set the values of 
the parameter $a_c$. The symbols are the results from the simulations, and the 
lines are the corresponding best-fit NFW profiles with the fitted concentration parameters listed in the plot.
The virial radius $R_{vir}$ is defined as the radius within which the average mass density of the halo
is $\Delta_{vir}\bar \rho_{DM}$. Here we take $\Delta_{vir}=337$ obtained for the $\Lambda$CDM
model with $\Omega_{DM0}=0.3$ \cite{Bullock01}. Studies have shown that both $\delta_c$ and $\Delta_{vir}$
are very insensitive to dark energy models (e.g, \cite{Pace10}).
It can be seen that the simulations with the setting described in this section
indeed give rise to dark matter halos with desired properties. 

To investigate the influence of different physical ingredients on the behavior of dark energy perturbations, 
we run simulations with different parameters as listed in Table~\ref{tab:sim}. 
F0 to F7 are simulations with the dark energy component described by an ideal fluid with the values of
$w$ and $c_s^2$ given in the second and third column. The $M_{vir}$, $R_{vir}$ and $c_{vir}$
are the virial mass, virial radius and the fitted NFW concentration parameter measured from 
simulations. The last two models are for the quintessence dark energy with the 
double exponential potential given by Eq.~(\ref{eq:poten}). The `SQ' run calculates the dark energy perturbations
by considering the scalar field evolution shown in Eq.~(\ref{eq:sf}), and the 
`FQ' run uses the fluid approach with the corresponding $w$ and $c_s^2$ to that of the quintessence model
(see Eq.~(\ref{eq:rhop}) and Eq.~(\ref{eq:wp})). 

\begin{table}
\caption{\label{tab:sim}Parameters of simulations}
\begin{ruledtabular}
\begin{tabular}{lcccccc}
    &$w$     &$c_s^2$&$a_c/a_0$&   $M_{vir}$    & $R_{vir}$& $c_{vir}$ \\
    &        &       &         &$(10^{15}M_{\odot})$ & (Mpc) &($R_{vir}/r_s$)\\
\hline
F0  & -0.9   &  0.5  &  0.8    & 1.00   & 2.6  & 5.7  \\ 
F1  & -0.8   &  0.5  &  0.7    & 1.06   & 2.6  & 5.1  \\ 
F2  & -0.7   &  0.5  &  0.67   & 1.08   & 2.7  & 5.6  \\ 
F3  & -0.9   &0.00001&  0.8    & 1.10   & 2.7  & 4.9  \\
F4  & -0.9   &  0.05 &  0.8    & 1.10   & 2.7  & 5.2  \\ 
F5  & -0.9   &   1   &  0.8    & 1.09   & 2.7  & 5.3  \\
F6  & -0.9   &  0.5  &  0.5    & 0.104  & 1.2  & 7.2  \\ 
F7  & -0.9   &  0.5  &  0.38   & 0.0082  & 0.52 & 9.9  \\ 
SQ  & 2exp   &   -   &  0.7    & 1.23   & 2.8  & 5.7  \\
FQ  & 2exp   &   -   &  0.7    & 1.25   & 2.8  & 5.7  \\ 
\end{tabular}
\end{ruledtabular}
\end{table}

\section{Results\label{sec:result}}
In this section, the results on dark energy perturbations and their dependence on 
different physical parameters are presented. The model F0 is taken to be the reference model. 

\subsection{Reference model}

We first show the temporal and spatial behaviors of dark energy perturbations for F0.

\begin{figure*}
\begin{tabular}{cc}
\resizebox{0.48\textwidth}{!}{\includegraphics{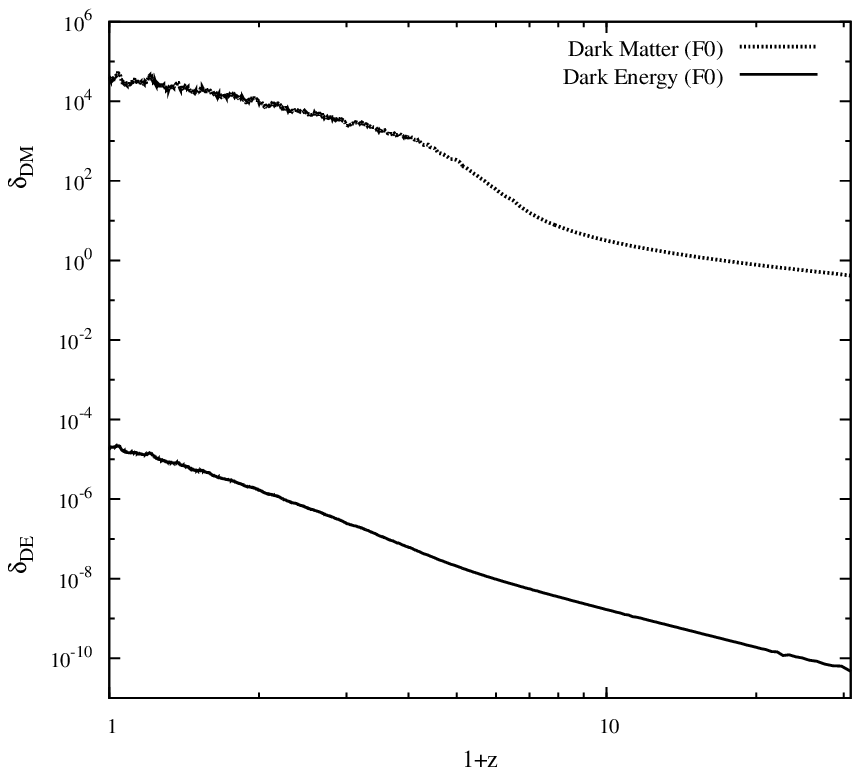}} &
\resizebox{0.48\textwidth}{!}{\includegraphics{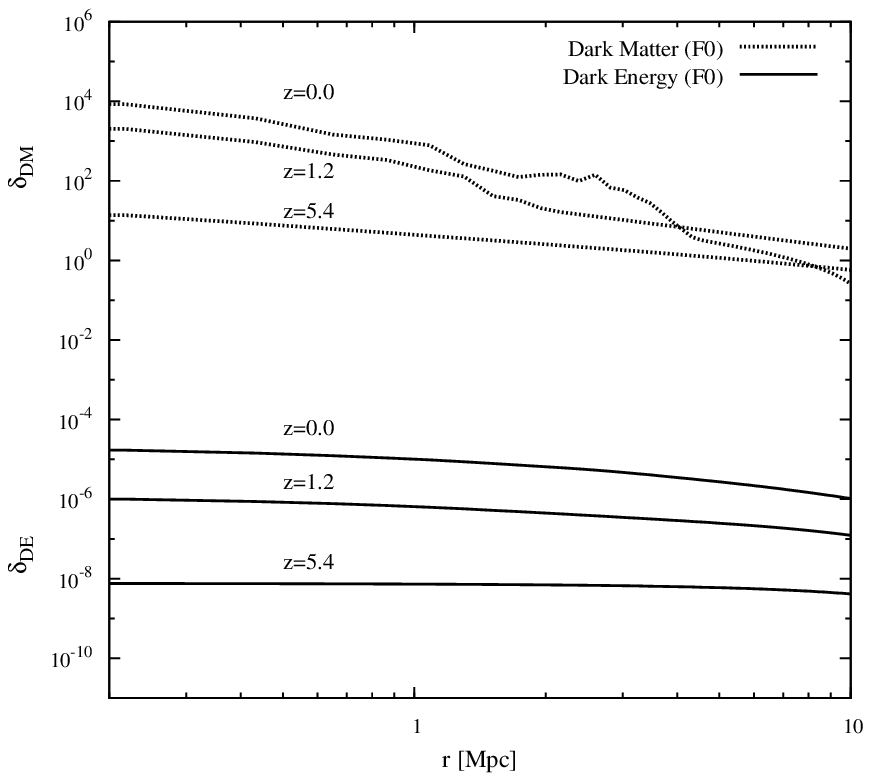}}
\end{tabular}
\caption{\label{fig:F0}
Left panel: The redshift evolution of the dark energy perturbation $\delta_{DE}$ (solid line)for the radial-mesh bin 
$r=0.1\hbox{ Mpc}$ is shown for reference model F0 ($M=10^{15} M_{\odot}, w=-0.9, c_s^2 =0.5$). The corresponding dark matter density 
perturbation $\delta_{DM}$ is also shown (dotted line). 
Right panel: The spatial profiles of $\delta_{DE}$ (solid lines) and $\delta_{DM}$ (dotted lines)
are shown for model F0 at $z=0$, $z=1.2$ and $z=5.4$, respectively.}
\end{figure*}

The left panel of Figure~\ref{fig:F0} shows the redshift evolution of the dark matter density contrast $\delta_{DM}$
and the dark energy perturbation $\delta_{DE}$ for the most inner radial-mesh bin $r=0.1\hbox{ Mpc}$. 
It can be seen that for the central part, the dark matter virialization happens at $z\sim 4$. 
The value of $\delta_{DM}$ reaches $\sim 10^5$ at $z=0$. For $\delta_{DE}$, it is always positive 
and increases smoothly without a sharp change corresponding to the 
virialization of dark matter at $z\sim 4$. Note that for F0, $c_s^2=0.5$, 
which gives rise to a resistance to the perturbation growth for the dark energy component
and explains its different time evolution behavior in comparison with that of the dark matter. 
At $z=0$, we have $\delta_{DE}\sim 10^{-5}$, which is about ten orders of magnitude smaller than $\delta_{DM}$.

The right panel of Figure~\ref{fig:F0} shows the spatial profiles for $\delta_{DM}$ and $\delta_{DE}$, respectively,
at different $z$. At $z>5$, the halo is in its early formation stage, and no dark matter virializations occur.
As the evolution proceeds, the virializations start from the central 
region and continually extend to larger regions. At $z=1.2$, the virialized part reaches
$r\sim 0.8\hbox{ Mpc}$ as indicated by the change of the $\delta_{DM}$ profile. 
At $z=0$, the virial region extends to $r \approx 2.6 \hbox{ Mpc}$, fully consistent with the 
virial radius defined by the average density contrast $\Delta_{vir}\approx 337$ with respect to the average
matter density of the universe for $\Omega_{DM0}=0.3$.  
For the dark energy perturbation $\delta_{DE}$, its amplitude grows with the decrease of the redshift,
and the profile gets steeper. As discussed above, unlike the behavior of $\delta_{DM}$ from which 
the virialized and unvirialized regions can be easily identified, $\delta_{DE}$ is rather smooth and there is not 
an apparent feature at the dark matter virialization boundary. Furthermore, the profile of $\delta_{DE}$ is
much shallower than that of the dark matter. From $r=0.2\hbox{ Mpc}$ to $r=10\hbox{ Mpc}$, $\delta_{DE}$ 
decreases from $10^{-5}$ to $10^{-6}$, whereas for $\delta_{DM}$, it changes from $\sim 10^{4}$ to $O(1)$.

\subsection{Dependences of $\delta_{DE}$ on $w$ and $c_s^2$}
\begin{figure}[ht]
\includegraphics[width=0.48\textwidth]{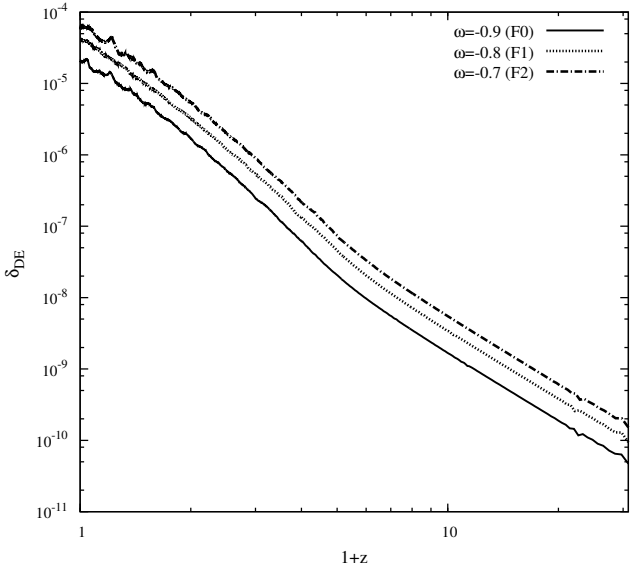}
\caption{\label{fig:F012}
The redshift evolution of $\delta_{DE}$ for the radial-mesh bin $r=0.1\hbox{ Mpc}$ is shown for F0 (solid line), F1 (dotted line) and F2 (dot-dashed line), respectively. }
\end{figure}

In Figure~\ref{fig:F012}, we show the dependence of $\delta_{DE}$ on the equation of state parameter $w$ of
the dark energy component. The time evolution of $\delta_{DE}$ for the inner-most radial-mesh bin  
is presented for $w=-0.9$ (F0), $w=-0.8$ (F1) and $w=-0.7$ (F2), respectively. 
The parameter $c_s^2$ is fixed to be $c_s^2=0.5$.  
The $w$-dependence is clearly seen. For $w=-0.7$, the dark energy perturbation $\delta_{DE}$ is
about three times as large as that with $w=-0.9$.

\begin{figure}[ht]
\includegraphics[width=0.48\textwidth]{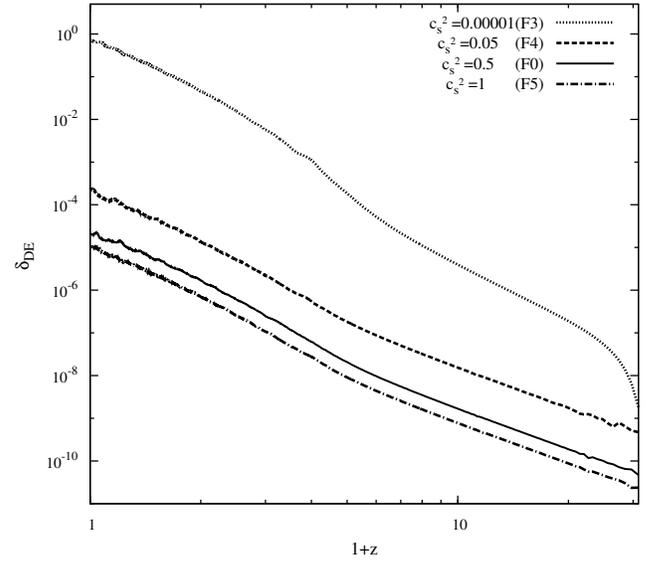}
\caption{\label{fig:F0345}
The $c_s^2$-dependence of $\delta_{DE}$. The redshift evolution of $\delta_{DE}$ 
for the inner-most bin at $r=0.1\hbox{ Mpc}$ is shown for F3 (dotted line),
 F4 (dashed line), F5 (dot-dashed line) and F0 (solid line). }
\end{figure}

Figure~\ref{fig:F0345} shows the effects of $c_s^2$ on dark energy perturbations
with $c_s^2=1.0$ (F5), $0.5$ (F0), $0.05$ (F4) and $0.00001$ (F3), respectively. 
A sensitive dependence of $\delta_{DE}$ on $c_s^2$ is apparent. For the
three cases with $c_s^2 \ge 0.05$, the dark energy perturbation amplitude
increases with time smoothly without characteristic features corresponding to the virialization of the dark matter
component. This is because the large sound speed of the dark energy can coordinate
its behavior over a large range quickly in comparison with that of the dark matter.
As expected, the amplitude of $\delta_{DE}$ is larger for smaller $c_s^2$.
For $c_s^2=0.05$, $\delta_{DE}$ is about an order of magnitude higher than that of
$c_s^2=0.5$. For a very small $c_s^2=0.00001$, we can see that at $z\sim 4$, 
there is a weak feature indicating the virialization of the dark matter component.
The amplitude of $\delta_{DE}$ in this case reaches $O(1)$ at $z=0$ in the central region.
In outer parts of the halo and at high redshifts, we still have $\delta_{DE}<1$. 
Thus our linear analyses for dark energy perturbations are still approximately valid for $c_s^2=0.00001$. 
For even lower $c_s^2$, we expect that the dark energy perturbation would follow that of the dark matter
more closely and can reach an amplitude of $\delta_{DE}>1$. In such cases, our formulation
presented in Sec. II for linear dark energy perturbations are not applicable, and 
nonlinear dark energy perturbations must be considered carefully. 
Moreover, the feedback effects of dark energy perturbations on the 
formation of the dark matter halo may also need to be taken into account (e.g., \cite{Creminell10}).   

\subsection{Dependence of $\delta_{DE}$ on properties of dark matter halos}

\begin{figure*}
\begin{tabular}{cc}
\resizebox{0.48\textwidth}{!}{\includegraphics{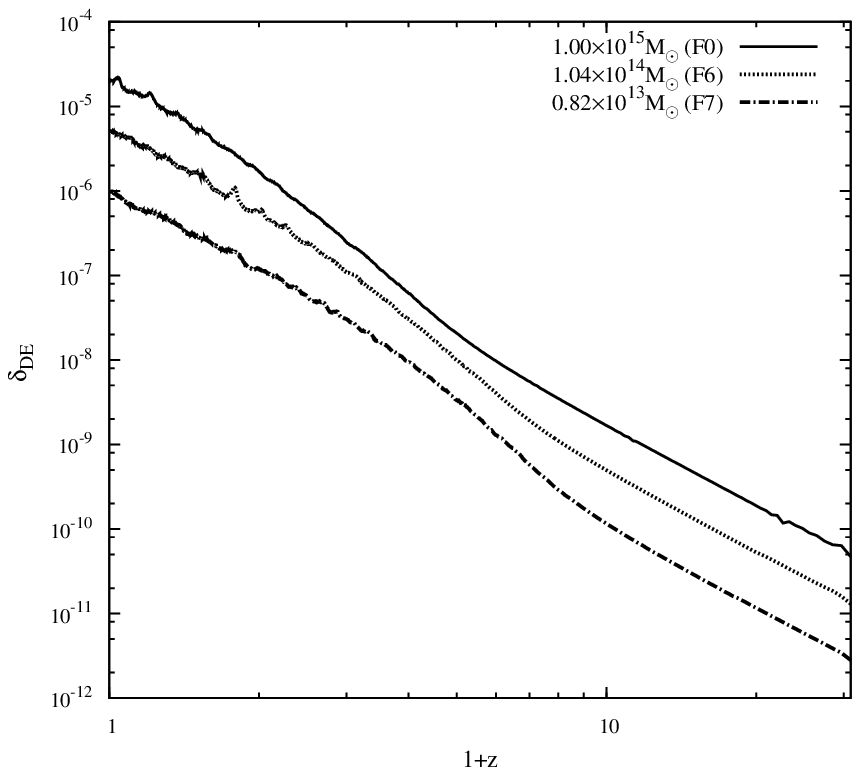}} &
\resizebox{0.48\textwidth}{!}{\includegraphics{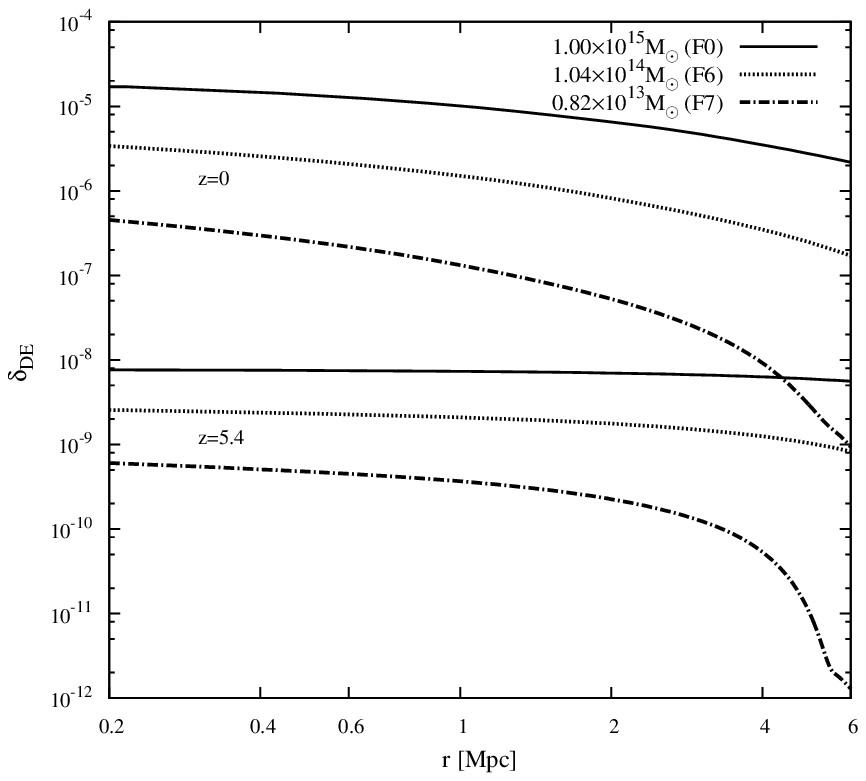}}
\end{tabular}
\caption{\label{fig:F067}
The left panel shows the dependence of $\delta_{DE}$ on the mass of the dark matter halo.
The evolution of $\delta_{DE}$ at $r=0.1\hbox{ Mpc}$ is shown
for F0, F6 and F7, respectively.  The right panel shows the spatial distribution of $\delta_{DE}$ for the three models at
$z=5.4$ (the lower set) and $z=0$ (the upper set), respectively.
}
\end{figure*}

Here we study the dark energy perturbations induced by different dark matter halos.
In the left panel of Figure~\ref{fig:F067}, the time evolution of $\delta_{DE}$
for the inner-most bin is shown for $M=1.00\times 10^{15}\hbox{ M}_{\odot}$ (F0), 
$1.04\times 10^{14}\hbox{ M}_{\odot}$ (F6), and $0.82\times 10^{13}\hbox{ M}_{\odot}$ (F7), respectively. 
It is seen that more massive halos induce larger dark energy perturbations. At $z=0$,
$\delta_{DE}$ of F0 is about $20$ times larger than that of model F7. 

The right panel of Figure~\ref{fig:F067} shows the spatial profiles of $\delta_{DE}$ for the three halos
at $z=5.4$ and $0$, respectively. It is noted that for smaller halos, the dark matter density profile is
more concentrated, and occupies smaller regions. Consequently, the profile for $\delta_{DE}$ 
is steeper and extends less for less massive halos. On the other hand, in all the three cases, the dark energy 
perturbations extend to regions much larger than the virial radii of the halos. 
The somewhat different line shapes in the left panel of Figure~\ref{fig:F067} are related
to the different virialization epochs for the central region for different models. The less massive halo
virializes earlier. 

\begin{figure}[ht]
\includegraphics[width=0.48\textwidth]{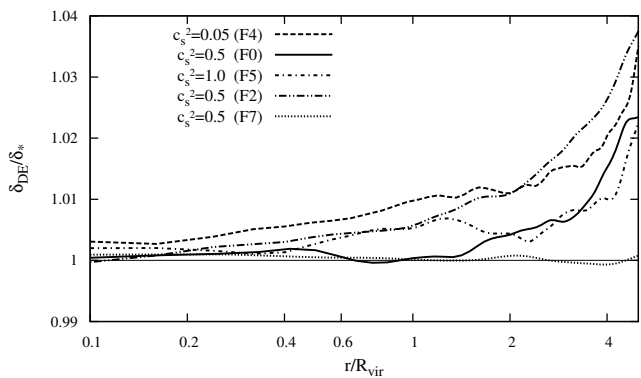}
\caption{\label{fig:dltc}
The validity of the relation given by Eq.~(\ref{eq:d2p}) between $\delta_{DE}$ and $\Phi$ is shown for different 
models. $\delta_{*} \equiv -(1+w)\Phi/c_s^2$. 
}

\end{figure}

We further analyze the approximate relation between $\delta_{DE}$ and the halo properties.
On halo scales, we have $k/a \gg H$. From the dark energy perturbation equations Eq.~(\ref{eq:dlt})-(\ref{eq:the}), 
we thus have approximately
\begin{equation}
\frac{c_s^2}{1+w}\delta_{DE} + \Phi \approx 0, \quad \delta_{DE} \approx -\frac{1+w}{c_s^2} \Phi.
\label{eq:d2p}
\end{equation}
In Figure~\ref{fig:dltc}, we show $\delta_{DE}/\delta_{*}$ at $z=0$ for a number of models
with different mass of dark matter halos, different $w$ and $c_s^2$ for the dark energy component, 
where $\delta_{*} \equiv -(1+w)\Phi/c_s^2$. 
It is seen that for all the models, $|\delta_{DE}/\delta_{*}-1|<1\%$ within the virial radius 
and $|\delta_{DE}/\delta_{*}-1|<4\%$ for $r<5R_{vir}$.
This demonstrates that the dark energy perturbation traces the gravitational potential of dark matter halos
very well over a broad range of model parameters, which may provide us an efficient way 
to estimate the power spectrum of dark energy perturbations in 
the regime of nonlinear structure formation. 
A similar tight correlation between $\delta_{DE}$ and $\Phi$ 
is also shown in \cite{BLi11a} for the extended quintessence model.

\subsection{Correspondence between the fluid approach and the scalar field model}

As discussed in Sec. II, the correspondence between a scalar field dark energy model and its fluid description
can be found by specifying suitable $w$ and $c_s^2$ that depend on the dynamical evolution of the scalar field.
Such correspondences have been applied extensively in analyzing the effect of the background dark energy 
on the structure formation and the behavior of the dark energy perturbation in the linear regime
of the structure formation (e.g., \cite{GZhao05}\cite{GZhao07}).
Here we test their validity in studying dark energy perturbations induced by the nonlinear
structure formation. We consider the scalar field model with a double exponential potential given
in Eq.~(\ref{eq:poten}) with $\alpha=6$ and $\beta=0.1$ , in accord with our previous studies \cite{Wang09}.
In the simulation runs SQ and FQ, we calculate the
dark energy perturbation from the scalar field perturbation given in Eq.~(\ref{eq:sf}) and from the fluid approach
with $w$ and $c_s^2$ changing with time according the dynamical evolution of the scalar field, respectively. 

\begin{figure}[ht]
\includegraphics[width=0.48\textwidth]{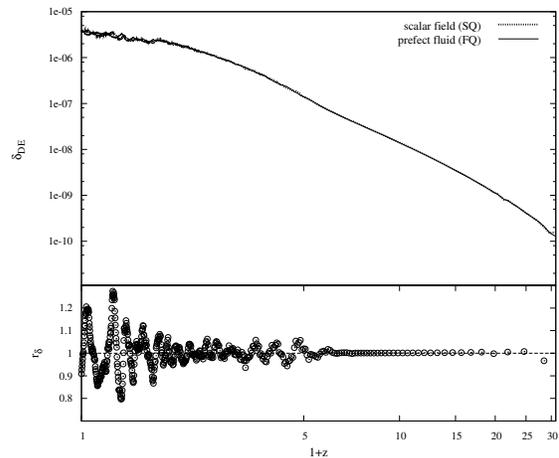}
\caption{\label{fig:SF}
The redshift dependence of $\delta_{DE}$ for the inner-most bin is shown in the upper panel for 
SQ and FQ, respectively. In the lower panel, the ratio $r_{\delta}$ of the two is shown.
}
\end{figure}

\begin{figure}[ht]
\includegraphics[width=0.48\textwidth]{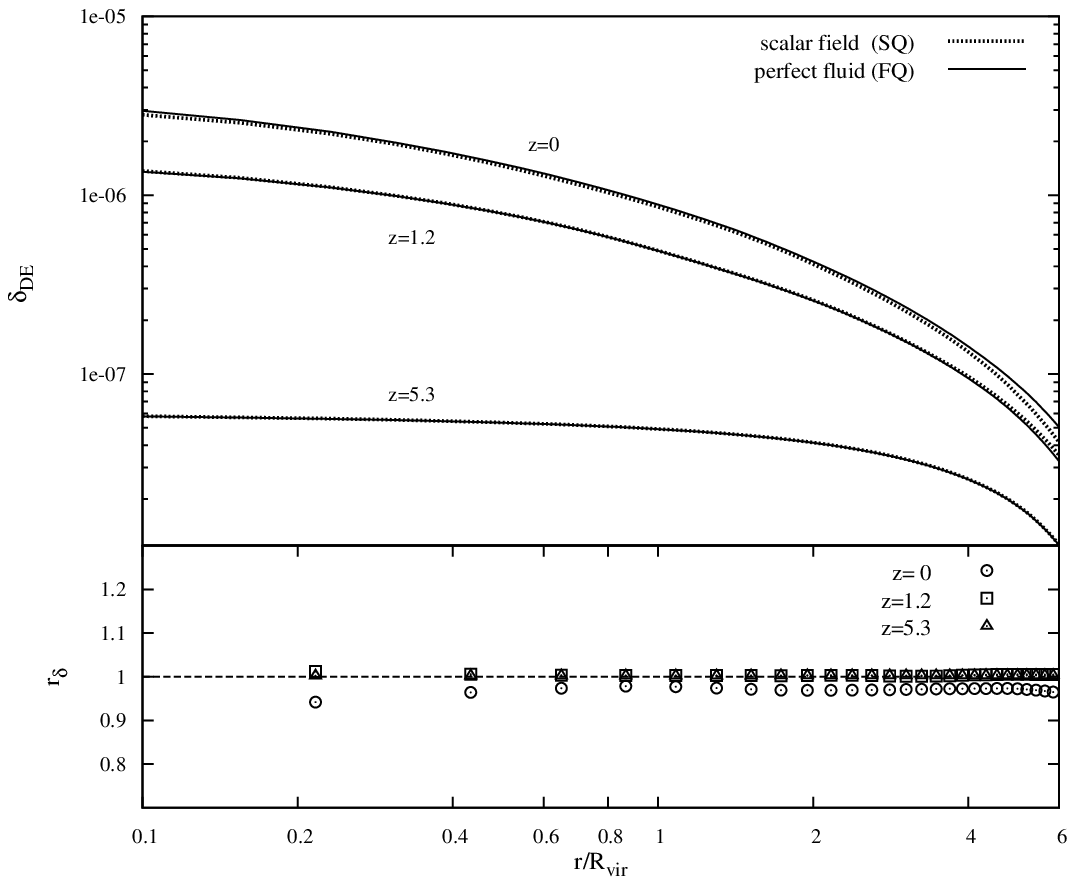}
\caption{\label{fig:SFp}
The spatial behaviors of $\delta_{DE}$ for SQ and FQ are shown in the upper panel for
$z=5.3$, $1.2$ and $z=0$, respectively. The corresponding $r_{\delta}$ are shown in the lower panel.
}
\end{figure}

The time evolution of $\delta_{DE}$ for the inner-most bin is shown in the upper panel of Figure~\ref{fig:SF}
for SQ and FQ, respectively. The ratio of the two $r_{\delta}=\delta_{DE}(SQ)/\delta_{DE}(FQ)$ is shown 
in the lower panel. The corresponding spatial behaviors at $z=5.3, 1.2$ and $z=0$ are shown in Figure~\ref{fig:SFp}.
It can be seen that the results from SQ and FQ match very well. The relative large scatter with
$r_{\delta}\sim 20\%$ at low redshift for the inner-most bin are due to the isotropized virialization that 
induces random dispersions for the gravitational potential somewhat differently in different model runs. 
These comparisons demonstrate that the fluid approach can be applied to study the dark energy perturbation behavior 
in the whole regime of structure formation, from linear to nonlinear stages.

\section{Summary and Discussion\label{sec:discussion}}

With two-component 1-D numerical simulations taking into account the fast and slow growth of dark matter halos,
we analyze the behavior of dark energy perturbations induced by the formation of spherical dark matter halos.
By comparing the results calculated directly from the scalar field evolution and that from 
the fluid description for a quintessence dark energy model with a double-exponential potential,
we show that the fluid approach can be used to analyze the dark energy behavior very well even in the nonlinear
stage of structure formation. In the fluid treatment for dark energy, the equation of state $w$ and the 
sound speed $c_s$ are the two important parameters that affect the dark energy perturbations significantly. 
Our studies find that in general, the dark energy perturbation arising from the formation of a dark matter halo
has a much more extended profile than that of the dark matter halo except for the case $c_s\approx 0$ where
the dark energy component follows the dark matter closely. A relation of $\delta_{DE}\approx [-(1+w)/c_s^2] \Phi$
with an accuracy about $1-2\%$ within the virial radius of a halo is revealed, 
which provides us a potential means to estimate the
power spectrum of dark energy perturbations from that of the dark matter potential field. 
From $w=-0.9$ to $w=-0.7$, $\delta_{DE}$ increases about three times from $\sim 2.5\times 10^{-5}$ to 
$7.5\times 10^{-5}$ for $c_s^2=0.5$. On the other hand, varying $c_s^2$ from $0.5$ to $0.05$,
the dark energy perturbation increases by an order of magnitude from $\sim 10^{-5}$ to $\sim 10^{-4}$. 
Our analyses show that our simulation treatment and the linear evolution of 
dark energy perturbations are valid for $c_s^2$ down to $c_s^2\sim 0.00001$.
For even smaller $c_s^2$, more accurate analyses taking into
account nonlinear dark energy perturbations and their feedback effects on the formation of dark matter halos
are needed (e.g.,\cite{Creminell10}).
The dependence of the dark energy perturbation
on the mass of the halo is also analyzed. It is shown that $\delta_{DE}$ increases by $\sim 20$ times
from $M\approx 10^{13}\hbox{ M}_{\odot}$ to $M\approx 10^{15}\hbox{ M}_{\odot}$.

The numerical analyses presented in this paper are done for dark energy models with $w >-1$. Our
simulations show that for such models, the dark energy perturbation induced by the formation of dark matter halos
has a clustering behavior with $\delta_{DE}>0$ during the entire evolutionary process.
This can also be seen clearly from the approximate relation $\delta_{DE}\approx [-(1+w)/c_s^2] \Phi$ (note $\Phi<0$
in dark matter halo regions). On the other hand, for $w< -1$, dark energy voids are expected from this relation
although we do not study these cases explicitly in our simulations. These results are in 
good agreement with studies shown in, e.g., \cite{Abramo07}\cite{Basse11}.
However, there are other analyses that point to the existence of dark energy void with $\delta_{DE}<0$
during the quasi-linear stages of dark matter halo formation even for models with $w> -1$
(e.g., \cite{Dutta07}\cite{Mota08}\cite{Wang09}). It is noticed that all the latter studies are performed using  
the Lemaitre-Tolman-Bondi (LTB) metric that is in the synchronous gauge. On the other hand, our simulation analyses presented here adopt 
the Newtonian gauge. It is well known that cosmological energy density perturbations are gauge dependent 
(e.g., \cite{Bardeen80}\cite{Ma95}\cite{Christopherson10}). Concerning the dark energy density perturbation in 
the LTB synchronous gauge $\delta_{DE}(Syn)$ and in the Newtonian gauge $\delta_{DE}(New)$, 
we have the relation \cite{Christopherson10}\cite{Ma95}
\begin{equation}
\delta_{DE}(Syn) = \delta_{DE}(New)-\alpha\frac{\dot {\bar {\rho}}_{DE}}{\bar {\rho}_{DE}},
\label{eq:syn_new_1}
\end{equation}
where $\alpha(t,\vec x)$ is related to the coordinate transformation for the time component associated with the 
two gauges, $t(Syn)=t(New)+\alpha$, and $\partial \alpha /\partial t=\Phi$.
We have ${\dot {\bar {\rho}}_{DE}}/{\bar {\rho}_{DE}}=-3(1+w)H$.
Further with the approximate relation between $\delta_{DE}(New)$ and $\Phi$, we then obtain 
\begin{equation}
\delta_{DE}(Syn) \approx \delta_{DE}(New)-3(1+w) H c_s^2 \int \frac{\delta_{DE}(New)}{1+w} dt.
\label{eq:syn_new_2}
\end{equation}
Thus if the second term is larger than the first term, a dark energy void with $\delta_{DE}(Syn)<0$ can occur
in the LTB synchronous gauge even for $w> -1$ with $\delta_{DE}(New)>0$. In other words, our results about the dark energy clustering for $w> -1$
presented in this paper do not conflict with those showing the existence of dark energy voids for the same models.
Instead, the differences merely reflect the differences of the specific gauge used in different analyses.

\begin{acknowledgments}
This research is supported in part by NSFC of China under grants 10533010, 10773001, 11033005,
and 973 program 2007CB815401.
\end{acknowledgments}

\bibliography{apssamp}

\end{document}